\documentclass[conference]{IEEEtran}
\IEEEoverridecommandlockouts
\usepackage{amsmath,amsfonts,amsthm,amssymb}
\usepackage{algorithmic}
\usepackage{algorithm}
\usepackage{array}
\usepackage{textcomp}
\usepackage{subfigure}
\usepackage{float}
\usepackage{enumerate}
\usepackage{tabularx}
\usepackage{stfloats}
\usepackage{url}
\usepackage{hyperref}
\usepackage{xcolor, soul,framed} 
\usepackage{xcolor}
\usepackage{verbatim}
\usepackage{graphicx}
\usepackage{epstopdf}
\usepackage{cite}
\usepackage{adjustbox}
\usepackage{pstricks}

\def\BibTeX{{\rm B\kern-.05em{\sc i\kern-.025em b}\kern-.08em
    T\kern-.1667em\lower.7ex\hbox{E}\kern-.125emX}}

\begin{document}
\newtheorem{Remark}{Remark}
\newtheorem{remark}{Remark}
\renewcommand{\algorithmicrequire}{\textbf{Input:}} 
\renewcommand{\algorithmicensure}{\textbf{Output:}}
\title{Optimal Discrete Beamforming of RIS-Aided Wireless Communications: An Inner Product Maximization Approach\\
\thanks{This work is supported by the National Natural Science Foundation of China (NSFC) under Grants NO.12141107, and the Interdisciplinary Research Program of HUST (2023JCYJ012).}}

\author{\IEEEauthorblockN{Rujing Xiong, Xuehui Dong, Tiebin Mi, Kai Wan, Robert Caiming Qiu}
\IEEEauthorblockA{School of Electronic Information and Communications, Huazhong University of Science and Technology\\Wuhan 430074, China. 
Email:\{rujing, xuehuidong, mitiebin, kai\_wan, caiming\}@hust.edu.cn}
}

\maketitle

\begin{abstract}

This paper studies the beamforming optimization challenge in reconfigurable intelligent surface (RIS)-aided multiple-input single-output (MISO) systems, where the RIS phase configuration is discrete. Conventional optimization methods for this discrete optimization problem necessitate resource-intensive exponential search and thus fall within the universal (NP-hard) category. We formally define this task as a discrete inner product maximization problem. Leveraging the inherent structure of this problem, we propose an efficient divide-and-sort (DaS) search algorithm to reach the global optimality for the maximization problem. The complexity of the proposed algorithm can be minimized to $\mathcal{O}(2^BN)$, a linear correlation with the count of phase discrete levels $2^B$ and reflecting units $N$. This is notably lower than the exhaustive search complexity of $\mathcal{O}(2^{BN})$. Numerical evaluations and experiments over real prototype also demonstrate the efficiency of the proposed DaS algorithm. Finally, by using the proposed algorithm, we show that over some resolution quantization level on each RIS unit (4-bit and above), there is no noticeable difference in power gains between continuous and discrete phase configurations.

\end{abstract}


\section{Introduction}

\IEEEPARstart{T}{he} reconfigurable intelligent surface (RIS) technique has recently demonstrated its great potential for reconfiguring wireless propagation environments~\cite{basar2019wireless}.
The advantage, as compared to other competitive technologies, lies in the fact that RISs provide opportunities for the so-called passive relays. RISs consist of a large number of carefully designed electromagnetic units and result in electromagnetic fields with controllable behaviors. It has recently been shown that RIS is a crucial enabler and presents a paradigm shift for future wireless networks. 

RISs have emerged as a promising solution for enhancing coverage by improving signal-to-noise ratio (SNR) and suppressing interference. To fully harness their advantages, RISs require precise and meticulous phase configurations. However, due to hardware limitations, most existing RIS prototypes employ low-resolution quantization scheme~\cite{xiong2023ris,arun2020rfocus, pei2021ris,dai2020reconfigurable,rains2021high}, such as the widely adopted 1-bit phase quantization scheme. 

Mathematically, the phase configuration optimization of RISs in practical beamforming is subject to two constraints. The first constraint is the unit modulus~\cite{wu2019intelligent} and the second constraint involves discrete phase configuration for practical implementations~\cite{wu2019beamforming}. In addressing the first constraint, various methods have been developed, including the alternating direction method of multipliers~\cite{wang2022sca,kumar2022novel}, semi-definite relaxation (SDR-SDP)~\cite{cui2019secure,zhou2020robust}, manifold optimization (Manopt)~\cite{yu2019miso, elmossallamy2021ris}. Furthermore, accommodating the second discrete configuration constraint, the most straightforward approach is to round the continuous phase configuration to the nearest point in the discrete constraint set~\cite{wang2020intelligent}. However, these two-stage approaches can lead to significant performance degradation~\cite{wu2019beamforming}, particularly when applied to extremely low-bit phase configurations. In the worst-case scenario, they may even result in arbitrarily poor performance~\cite{zhang2022configuring}.

Other approaches directly handle the discrete phase configurations. It is essential to recognize that discrete optimization problems tend to fall within the universal (NP-hard) category, requiring resource-intensive exponential search techniques. In response to this challenge, researchers have put forward various techniques aimed at accelerating the search process in~\cite{di2020hybrid,zhang2022configuring}. These methods significantly reduce search complexity but often converge to local optima. A rotation-based algorithm has been proposed for global optimum in~\cite{ren2022linear}. However, due to the infinite rotation direction in high-dimensional space, this method is limited to single-input single-out (SISO) systems and is hard to extend to multiple-input single-output (MISO) systems. 

%
%

To the best of our knowledge, a significant research gap exists concerning the development of highly efficient search algorithms capable of ensuring global optimality for discrete phase configurations in MISO communications. To tackle this challenge, we redefine the signal enhancement problems as discrete inner product maximizations and propose a highly efficient divide-and-sort (DaS) search framework which guarantees the attainment of global optima.

\paragraph*{Contributions}
%
%


We propose a framework of discrete inner product maximization to generalize and address the beamforming problem in RIS-aided signal enhancement. This method is applicable to MISO systems and encompasses the special case of SISO systems. In this framework, the beamforming problem is initially formulated as a discrete inner product maximization problem. Subsequently, a mathematically concise search algorithm, named divide-and-sort (DaS), is introduced to solve the problem. The DaS algorithm is guaranteed to find the global optimum with polynomial search complexity. We verify the effectiveness of the proposed algorithm through extensive numerical simulations and experimental trials. Additionally, we demonstrate that discrete phase configurations exhibit similar performance to continuous configurations for moderate-resolution quantization (e.g., 4-bit) in RIS-aided communication systems.

\begin{figure}[tbp]
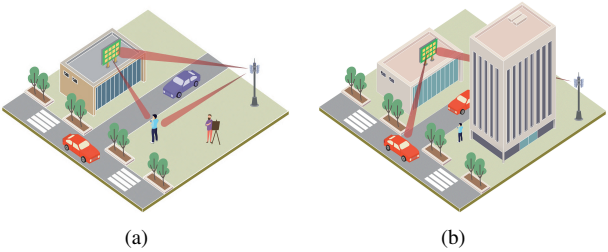

  \centering
  \subfigure[]{
  \label{With LoS}
  \includegraphics[width=.45\columnwidth]{Scenario2-1.eps}}
  \subfigure[]{
  \label{With NLoS}
  \includegraphics[width=.45\columnwidth]{Scenario1-1.eps}}
  \caption{RIS-aided point-to-point MISO communication. (a) With direct link. (b) Without direct link.}
  \label{Scenarios}
\vspace{-0.4cm}
\end{figure}

\paragraph*{Notations}
The imaginary unit is denoted by~$j$. The magnitude, real and complex component of a complex number are represented by~$|\cdot|$, $\mathcal{\Re}\{\cdot\}$ and $\mathcal{\Im}\{\cdot\}$ respectively. $\|\cdot\|$ denotes the Euclidean norm of a complex vector. Unless explicitly specified, lower and upper case bold letters denote vectors and matrices. The conjugate transpose and transpose of~$\mathbf{A}$ are denoted by~$\mathbf{A}^H$ and~$\mathbf{A}^T$, respectively. 

\section{System model and problem formulation}\label{Section2}

\addtolength{\topmargin}{0.04in}

\subsection{System Model}
We consider a point-to-point MISO downlink communication system where a base station (BS) equipped with $M$ antennas serves a single-antenna user equipment (UE). To enhance the communication quality between the BS and the UE, a RIS consisting of $N$ reflecting units is employed, with $M \ll N$. Let $x \in \mathbb{C}$ be the transmitted signal. If we denote the phase configuration vector related to the RIS by $\mathbf{w} =  [e^{j\omega_1},e^{j\omega_2},\dots, e^{j\omega_N}]^H$, where $\omega_i \in [0, 2 \pi), i = 1,\cdots, N$, the received signal at the UE can be expressed as
\cite{wu2019intelligent}
\begin{equation}
y = (\mathbf{h}_r^H\mathbf{W}\mathbf{G}+\mathbf{h}_d^H)\mathbf{s}x +v,
\end{equation}
where $\mathbf{W} = {\rm diag} (\mathbf{w})$, $\mathbf{s} \in \mathbb{C}^M$ denotes the transmit beamforming vector. The equivalent channel between BS-RIS and RIS-UE links are denoted by $\mathbf{G} \in \mathbb{C}^{N \times M}$ and $ \mathbf{h}_r^H\in \mathbb{C}^{N}$, respectively. The direct link direct channel between the BS and the UE is represented by $\mathbf{h}_d\in \mathbb{C}^M$. The received signal is corrupted by additive Gaussian white noise $v \sim \mathcal{CN}(0,\sigma^2)$.

\subsection{Problem Formulation}
The objective of RIS-aided signal enhancement is to maximize the received signal power, which can be formulated as
\begin{equation}\label{P1}
\begin{aligned}
 \max_{\omega_i, \dots, \omega_{N} \in [0, 2 \pi)}& \  |(\mathbf{h}_r^H\mathbf{W}\mathbf{G}+\mathbf{h}_d^H)\mathbf{s}|^{2}.\\
\end{aligned}
\end{equation}
By substituting optimal transmit beamforming solution $\boldsymbol{w}^*=\sqrt{P} \frac{\left(\boldsymbol{h}_r^H \mathbf{W} \mathbf{G}+\boldsymbol{h}_d^H\right)^H}{\left\|\boldsymbol{h}_r^H \mathbf{W} \mathbf{G}+\boldsymbol{h}_d^H\right\|}$\cite{wu2019intelligent, tse2005fundamentals}, 
\eqref{P1} is equivalent to
\begin{equation}\label{P2}
\begin{aligned}
 \max_{\omega_i, \dots, \omega_{N} \in [0, 2 \pi)}& \  \|(\mathbf{h}_r^H\mathbf{W}\mathbf{G}+\mathbf{h}_d^H)\|^{2}.\\
\end{aligned}
\end{equation}
In practical implementations, the phase configuration is selected from a finite set of values. To facilitate this, we denote the discrete configuration set by $\mathcal{U} = \left\{ 0, \Omega, \ldots, (2^B-1) \Omega \right\}$, where $B$ represents the phase quantization level and $\Omega= 2 \pi / 2^B$. With this quantization scheme, the discrete beamforming optimization problem can be formulated as
\begin{equation}\label{P3}
\begin{aligned}
 \max_{\omega_i, \dots, \omega_{N} \in \mathcal{U}}& \  \|(\mathbf{h}_r^H\mathbf{W}\mathbf{G}+\mathbf{h}_d^H)\|^{2}.\\
\end{aligned}
\end{equation}
Let $\mathbf{h}_r^H\mathbf{W}\mathbf{G} = \mathbf{w}^H\boldsymbol{\Phi}$ where $\mathbf{w}=\left[e^{j \omega_1}, e^{j \omega_2}, \dots, e^{j \omega_N} \right]^H$, and introduce augmented $\mathbf{\tilde{w}}=\left[e^{j \omega_1}, e^{j \omega_2}, \dots, e^{j \omega_N}, 1 \right]^H$, matrix $ \boldsymbol{ \tilde{ \Phi} } = [ \boldsymbol{\Phi}^T, \mathbf{h}_d ]^T \in \mathbb{C}^{(N+1)\times M}$.~\eqref{P3} can be written as 
\begin{equation}\label{P4}
\begin{aligned}
\text{(P1)} \ \max_{\omega_i, \dots, \omega_{N} \in \mathcal{U}}& \ \|\mathbf{\tilde{w}}^H\boldsymbol{\tilde{ \Phi }}\|.
\end{aligned}
\end{equation}
Note that $\boldsymbol{\tilde{ \Phi }}$ is with rank $M$ with high probability. 
Upon acquiring an solution $\mathbf{\tilde{w}}_{ \text{opt} }$ to problem ({P1}), the optimal solution for the original problem \eqref{P3} can be derived as $\mathbf{w}_\text{opt} = \mathbf{ \tilde{w}}_{\text{opt}} (1:N) /  \mathbf{\tilde{w}}_{\text{opt} } (N+1)$.

In certain exceptional scenarios, as depicted in Fig.~\ref{With NLoS}, the line of sight link between the BS and UE may blocked by obstacles. $\mathbf{h}_d^H$ are assumed to be $\mathbf{0}$ in such condition. Consequently, the optimization problem of maximizing the received signal power can usually be formulated as
\begin{equation}\label{P5}
\text{(P2)} \ \max_{\omega_i, \dots, \omega_{N} \in \mathcal{U}} \ \|\mathbf{w}^H\boldsymbol{\Phi}\|,
\end{equation}
where $\boldsymbol{\Phi}\in \mathbb{C}^{N\times M}$ is with rank $M$.

The optimization problem (P2) is a non-convex discrete multi-objective programming problem, which is NP-hard and has never been addressed. In this problem, matrix $\boldsymbol{\Phi}$ is low-rank. One approach to solving the problem is to perform an exhaustive search among all elements in $\mathcal{U}^{N}$, which has a computational complexity of $\mathcal{O}(2^{BN})$, making it intractable.

\textbf{Special Case:} single-input single-output (SISO). The above framework problem formulation involves the special case in which the number of transmitter antennas is 1. In this situation, the equivalent channel between the RIS-UE link can be denoted as $\mathbf{g} \in \mathbb{C}^N$. The optimization problem can be formulated as
\begin{equation}\label{ip}
\text{(P3)} \ \max_{\omega_i, \dots, \omega_{N} \in \mathcal{U}} \ |\mathbf{w}^H\boldsymbol{\phi}|,
\end{equation}
It is an inner product maximization problem, where $\boldsymbol{\phi} = {\rm diag} (\mathbf{h}_r^H \mathbf{g}) \in \mathbb{C}^N$ is with rank-1.

\section{A divide-and-sort search framework for discrete inner product maximization}\label{Section3}
Since the only difference between the problems (P1) and (P2) is in the numbers of optimization variables (which are $N+1$ and $N$, respectively), the main steps of the proposed optimization algorithms for these two problems are essentially the same. Hence, in this section we proposed the generalized divide-and-sort (DaS) algorithm to address the maximization problem (P2). Our proposed algorithm allows us to construct a parameterized search set with a cardinality of $(2^BN)^{2M-1}$, while the complexity of the exhaustive search is  $\mathcal{O}(2^{BN})$. 

This is practically advantageous, given that the quantity of RIS units $N$ is typically more than hundreds, while the number of transmitting antennas $M$ and phase quantization level $B$ are relatively limited. In the system with only a single transmit antenna, the search complexity can be reduced to $\mathcal{O}(2^BN)$. 

%

\subsection{Generalized Algorithm (low-rank)}
To address the problem (P2), we initially reformulate it into an inner product maximization problem. As $\boldsymbol{\Phi}$ is with rank $M$ (low-rank) and $\mathbf{w}^H\boldsymbol{\Phi}$ is a row vector of length $M$ in \eqref{P5}, we can introduce an auxiliary vector $\mathbf{v}(\boldsymbol{\theta},\boldsymbol{\psi}) = \left[e^{j \psi_1}\sin \theta_1, e^{j \psi_2}\cos\theta_1\cos\theta_2, \dots, e^{j \psi_M}\cos\theta_1 \ldots \cos\theta_{M-1}\right]^H$ to rephrase the optimization problem (P2) as 
\begin{equation}
\begin{aligned}
\max_{\omega_1, \ldots, \omega_{N} \in \mathcal{U}}  \max_{\boldsymbol{\theta},\boldsymbol{\psi}} \ \Re \left \{  \mathbf{v}(\boldsymbol{\theta},\boldsymbol{\psi})^H \boldsymbol{\Phi}^H \mathbf{w}  \right \},
\end{aligned}
\end{equation}
where $\boldsymbol{\theta},\boldsymbol{\psi} \in (0,2\pi]^N$, $\mathbf{v}(\boldsymbol{\theta},\boldsymbol{\psi}) \in \mathbb{C}^M$.

Let vector $\mathbf{a}^H = \mathbf{v}(\boldsymbol{\theta},\boldsymbol{\psi})^H\boldsymbol{\Phi}^H \in \mathbb{C}^{N}$ and $a_n^H = | a_n |e^{j\tau_n}$ denote the $n$-th element in $\mathbf{a}^H$, it can be written as 
\begin{equation}
\begin{aligned}
\text{(P4)} &\max_{\omega_1, \ldots, \omega_{N} \in \mathcal{U}}  \max_{\boldsymbol{\theta},\boldsymbol{\psi}} \ \Re \left \{  \mathbf{a}^H \mathbf{w}  \right \} \\ =
&\max_{ \omega_1, \dots, \omega_{N} \in \mathcal{U} } \max_{\boldsymbol{\theta},\boldsymbol{\psi}}  \Re \left \{ \sum_{n=1}^{N} |a_n| e^{j (\tau_n-\omega_n ) } \right \} \\ = 
&\max_{ \omega_1, \dots, \omega_{N} \in \mathcal{U} } \max_{\boldsymbol{\theta},\boldsymbol{\psi}}  \Re \left \{ \sum_{n=1}^{N} |a_n|  \cos(\tau_n-\omega_n)  \right \}.
\end{aligned} 
\end{equation}

The problem (P4) is an \textbf{inner product maximization problem}. Given a vector $\mathbf{v}(\boldsymbol{\theta},\boldsymbol{\psi})$, the maximum is achieved by selecting an appropriate $\omega_n = \omega_{n,\text{opt}}$ from the discrete set $\mathcal{U}$ to minimize the expression $\tau_n-\omega_n$, for $n = 1,2,\cdots, N$. In essence, we can divide the interval $(0,2\pi]$ into $N$ regions based on the discrete values int set $\mathcal{U}$, with the $n$-th region centered around $(n-1)\Omega$. Consequently, determining to which segment the observed $\tau_n$ belongs allows us to ascertain the optimal $\omega_{n,\text{opt}}$. For instance, while $\Omega/2<\tau_n<3\Omega/2$, $\omega_{n,\text{opt}}(\tau_n) = \Omega$, and 
$3\Omega/2<\tau_n<5\Omega/2$,  $\omega_{n,\text{opt}}(\tau_n) = 2\Omega$.

The angles $\tau_1,\cdots,\tau_N$ are determined by unknown vector $\mathbf{v}(\boldsymbol{\theta},\boldsymbol{\psi})$. It is important to highlight that the feasible region for $\mathbf{v}(\boldsymbol{\theta},\boldsymbol{\psi})$ is continuous and corresponds to a space $\mathcal{S}$ with dimensions $2M-1$. The continuity allows us to select an appropriate vector $\mathbf{v}(\boldsymbol{\theta},\boldsymbol{\psi})$ in $\mathcal{S}$ such that $\tau_n = \omega_n$. Similarly, we can utilize the discreteness of $\omega_n$ to determine boundaries in space $\mathcal{S}$ and partition the entire high-dimensional space into different subspaces. Each of these subspaces corresponds to a candidate of $\omega_1,\cdots, \omega_N$. By traversing candidates considering each subspace and its boundaries, the optimal solution $\boldsymbol{\omega}_{\text{opt}} = \left[ \omega_{1, \text{opt}}, \omega_{2, \text{opt}}, \cdots, \omega_{N, \text{opt}} \right]^T$ can be obtained.

We proceed to construct the feasible search set $\mathbb{U}$ comprising all candidates. For any $\alpha$ belonging to the set $\left\{ 0, \Omega, \ldots, (2^{B-1}-1) \Omega \right\}$\footnote{\label{foot:set} Due to the presence of symmetry, omitting the entire $\mathcal{U}$ does not affect the solutions of the boundary equations and establishment of the search set $\mathbb{U}$.},
we define the boundary conditions for the subspaces within $\mathcal{S}$ as follows:
\begin{equation}\label{selection}
a_n = \boldsymbol{\Phi}_{n,:}  \cdot  \mathbf{v}(\boldsymbol{\theta},\boldsymbol{\psi})  = ce^{j\alpha}, n =1,\cdots,N,
\end{equation}
where $\boldsymbol{\Phi}_{n,:} $ denotes the $n$-row of $\boldsymbol{\Phi}$,  and $c \in \mathbb{R}$. 

Upon choosing and combining $2M-1$ equations from the system~\eqref{selection}, an intersection point $\mathbf{v}(\boldsymbol{\theta},\boldsymbol{\psi})$ with $2M-1$ unknown variables can be determined. For instance:
\begin{equation}\label{bound}
\left[\begin{array}{c}
\boldsymbol{\Phi}_{1,:} \\
\boldsymbol{\Phi}_{2,:} \\
\vdots \\
\boldsymbol{\Phi}_{2M-1,:}
\end{array}\right] \cdot \mathbf{v}(\boldsymbol{\theta},\boldsymbol{\psi})=\left[\begin{array}{c}
c_1 e^{j \alpha_1} \\
c_2 e^{j \alpha_2} \\
\vdots \\
c_{2M-1} e^{j \alpha_{2M-1}}
\end{array}\right]
\end{equation}

Let $\mathbf{C}=\left[e^{-j \omega_1} \boldsymbol{\Phi}_{1,:}, e^{-j \alpha_2} \boldsymbol{\Phi}_{2,:},\cdots, e^{-j \alpha_{2 M-1}} \boldsymbol{\Phi}_{2M-1,:}\right]^T $. Consequently, the intersection point $\mathbf{v}(\boldsymbol{\theta},\boldsymbol{\psi})$ can be defined by solving $[\Re(\mathbf{C}) \quad \Im(\mathbf{C})]\left[\begin{array}{c}
\Im\left(\mathbf{v}(\boldsymbol{\theta},\boldsymbol{\psi})\right) \\
\Re\left(\mathbf{v}(\boldsymbol{\theta},\boldsymbol{\psi})\right)
\end{array}\right]=\mathbf{0}$. Each equation in~\eqref{bound} represents a boundary in $\mathcal{S}$, and $\mathbf{v}(\boldsymbol{\theta},\boldsymbol{\psi})$ is the point associated with $2^{2M-1}$ subspaces, thus related to $2^{2M-1}$ candidate solution.

Next, we will show how to determine the $2^{2M-1}$ candidates according to the single intersection. We substitute the obtained $\mathbf{v}(\boldsymbol{\theta},\boldsymbol{\psi})$ back into equation $\mathbf{a}^H = \mathbf{v}(\boldsymbol{\theta},\boldsymbol{\psi})^H\boldsymbol{\Phi}^H $ and $\tau_n^{\text{ori}} = \arg a_n^H $. Recall that $\omega^{\text{ ori}}_n = \tau^{\text{ ori}}_n$ is solution to problem (P4), a basic vector $\boldsymbol{\omega}^{\text {ori}} = [e^{j\tau^{\text{ ori}}_1},\cdots,e^{j\tau^{\text{ ori}}_N} ]^T$ can be defined. As the selected vector in~\eqref{bound} with index $n \ ( n = 1,\cdots, 2M-1)$ satisfying the boundary conditions, the element with index $n $ in the candidate $\boldsymbol{\omega}$ has two possible valves, $\tau^{\text{ ori}}_n$ and $\tau^{\text{ ori}}_n+\Omega$. To prevent the omission, we encompass both conceivable scenarios and thus recognize $2^{2M-1}$ candidate solutions. 


By traversing all possible equation combinations in~\eqref{selection}, we can select and solve $C_N^{2M-1}(2^{B-1})^{2M-1}$ different intersection points. Thus we can formulate a search set $\mathbb{U}$ consisting total of $C_N^{2M-1}(2^{B})^{2M-1}$ candidate solutions. The optimal solution to the problem (P2) can be achieved through traversing $\mathbb{U}$. we summarize the proposed algorithm in Algorithm\ref{alg1}.

\begin{algorithm}
\caption{Generalized DaS algorithm for low-rank} 
\label{alg1} 
\begin{algorithmic}[1] 
\REQUIRE Complex matrix $\boldsymbol{\Phi}$, the phase quantization level $B$. 
\ENSURE Optimal phase configurations  $\omega_{1,\text{opt}}, \ldots, \omega_{N,\text{opt}}.$ 
\STATE Determine and denote the rank of matrix $\boldsymbol{\Phi}$ as $M$. 
\REPEAT
\STATE Select $2M-1$ row vector in matrix $\boldsymbol{\Phi}$, record the indexes, and utilize them to formulate an equation group, as in~\eqref{bound}.
\STATE Solve the equations for the intersection $\mathbf{v}(\boldsymbol{\theta},\boldsymbol{\psi})$.
\STATE Substitute the $\mathbf{v}(\boldsymbol{\theta},\boldsymbol{\psi})$ into $\mathbf{a}^H = \mathbf{v}(\boldsymbol{\theta},\boldsymbol{\psi})^H\boldsymbol{\Phi}^H \in \mathbb{C}^{N}$, calculate $\{\tau_1,\cdots,\tau_N\} = \arg (\mathbf{a}^H)$ and basic vector $\boldsymbol{\omega}^{\text {ori}} = [e^{j\tau_1},\cdots,e^{j\tau_N} ]$.
\STATE Determine the elements in basic vector $\boldsymbol{\omega}^{\text {ori}}$, has undetermined value $\tau_n-\Omega$ or $\tau_n+\Omega$, collect all the $2^{2M-1}$ possible combinations as basic set $\{\boldsymbol{\omega}^{\text {ori}}_1,\cdots,\boldsymbol{\omega}^{\text {ori}}_{2^{2M-1}}\}$.  
\UNTIL {All possible $C_N^{2M-1}(2^{B-1})^{2M-1}$ equation combinations in\eqref{selection} are selected.}
\STATE Build the search set $\mathbb{U}$ by combining all the basic sets.
\STATE Find $\omega_{1, \text{opt}}, \ldots, \omega_{N, \text{opt}} = \arg \max_{ \{ \omega_1, \dots, \omega_{N} \} \in \mathbb{U} } \|\mathbf{w}^H\boldsymbol{\Phi}\|$.
\RETURN $\mathbf{w}_{\text{opt}}= \left[ e^{j \omega_{1, \text{opt}} }, \cdots, e^{j \omega_{N, \text{opt}} } \right]^H$.
\end{algorithmic} 
\end{algorithm}

%

\begin{figure}[htbp]
  \centering
  \includegraphics[width=1\linewidth]{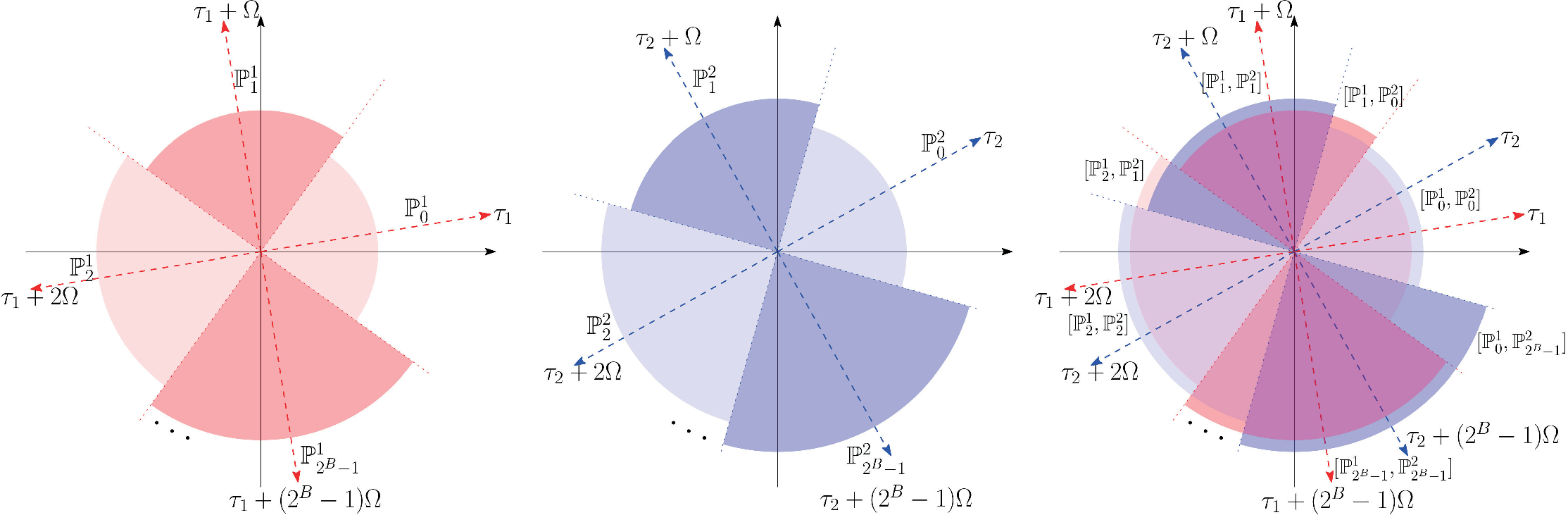}
  \caption{Illustration of region rearrangement in the case of rank-1 in inner product maximization problem. There are two groups of partitions ($N=2$), represented by red (left) and blue arcs (middle), each containing $2^B$ regions. After the rearrangement, these combine to form a total of $2^B \times 2$ non-overlapping regions (right).}
  \label{Fig:MultipleElements}
\vspace{-0.2cm}
\end{figure}

\subsection{Special Case Algorithm (rank-1)}
To enhance comprehension of the proposed algorithm, we simplified it to address the problem (P3), where $\boldsymbol{\phi}$ is with rank 1. The simplified approach entails a complexity of $\mathcal{O}(2^BN)$. 

A remarkable feature of the optimization problem (P3) is that $\mathbf{w} ^H \boldsymbol{\phi}$ is a scalar. Thus, we can introduce an auxiliary variable $\psi \in [0, 2 \pi)$ and rewrite the objective function as
\begin{equation}\label{Equivalent}
\begin{aligned}
& \max_{ \omega_1, \dots, \omega_{N} \in \mathcal{U} } \left| \mathbf{w} ^H \boldsymbol{\phi} \right| \\
= 
& \max_{ \omega_1, \dots, \omega_{N} \in \mathcal{U} } \max_{\psi \in [0, 2 \pi)}  \Re \left \{ \sum_{i=1}^{N} |\phi_n| e^{j ( - \psi + \tau_n + \omega_n ) } \right \} \\
= 
& \max_{\psi \in [0, 2 \pi)} \sum_{n=1}^{N} \left(|\phi_n| \max_{ \omega_n \in \mathcal{U} }  \cos \left( \psi - (\tau_n + \omega_n) \right) \right),
\end{aligned}
\end{equation}


The introduction of the auxiliary variable $\psi$ in \eqref{Equivalent} provides a significant advantage by decoupling the original problem. We next develop a practical method to solve the subproblem
\begin{equation}\label{E:Subproblem}
  \omega_{n, \text{opt}} (\psi) = \arg \max_{ \omega_n \in \mathcal{U} } |\phi_n| \cos \left( \psi - (\tau_n + \omega_n) \right).
\end{equation}
For each $1 \le n \le N$, $|\phi_n|$ and $\tau_n$ are known \emph{a priori}. We note that when $\psi$ is specified, 
the primary task to solve~\eqref{E:Subproblem} is to select $\omega_n$ in a way that minimizes $| \psi - (\tau_n + \omega_n) |$.

It is notable that for each $\tau_n$, the number of potential combinations $(\tau_n+\omega_n)$ is limited to $2^B$, expressed as $\tau_n + i\Omega, i = 0,1,\cdots,{2^B-1}$. We partition the interval $[0, 2 \pi)$ (a complete angle) into $2^B$ equal-length regions $\mathbb{P}^n_0,\mathbb{P}^n_1, \dots, \mathbb{P}^n_{2^B-1}$. Each region $\mathbb{P}^n_{i}= \left[ ( \tau_n + i \Omega ) - \Omega / 2, ( \tau_n + i \Omega ) + \Omega / 2 \right)$ is centered around $\tau_n + i\Omega$. 

When a $\psi$ is given, by checking which region $\psi$ belongs to, we can easily determine the optimal $\omega_{n, \text{opt}} (\psi)$. In other words, if $ ( \tau_n + i \Omega ) - \Omega / 2 \le \psi < ( \tau_n + i \Omega ) + \Omega / 2 $, then $\omega_{n, \text{opt}} (\psi) = i \Omega$ is determined. 
For each $n = 1,\cdots,N$, we define the solution vector as 
  $\boldsymbol{\omega}_{\text{opt}} (\psi) = \left[ \omega_{1, \text{opt}} (\psi) , \omega_{2, \text{opt}} (\psi), \cdots, \omega_{N, \text{opt}} (\psi) \right]^T$, 
$\boldsymbol{\omega}_{\text{opt}} (\psi)$ is a piecewise constant function and takes at most $2^B N$ distinct values as $\psi$ varies across the interval $[0, 2\pi)$. We gather all the $2^B N$ candidates to form the search set $\mathbb{U}$. 

To achieve this, we initially denote the centers associated with $\tau_n$ as $\mathcal{C}^n = \{ \tau_n, \tau_n + \Omega, \cdots, \tau_n + (2^B-1) \Omega \} \mod 2 \pi$, $n = 1, \cdots, N$. Without loss of generality, the smallest element in $\mathcal{C}^n$ is assumed to be $\tau_n$, where $0 \le \tau_n < \Omega$ for each $n$. 



Next, we establish a bijective mapping $\mathcal{M}$ to sort $\{ \tau_1, \ldots, \tau_N \}$ such that $\tau_{\mathcal{M}(1)} \le \tau_{\mathcal{M}(2)} \le \cdots \le \tau_{\mathcal{M}(N)}$.
For each region $\mathbb{P}$, the basic vector can be expressed as
\begin{equation*}
\begin{aligned}
\mathbb{P}^1 & : [ \tau_{\mathcal{M}(1)} + \Omega, \tau_{\mathcal{M}(2)}, \cdots, \tau_{\mathcal{M}(N)} ]^T , \\
\vspace{-5mm}
\vdots \\
\mathbb{P}^{N} & : [ \tau_{\mathcal{M}(1)} + \Omega, \tau_{\mathcal{M}(2)} + \Omega, \cdots, \tau_{\mathcal{M}(N)} + \Omega ]^T, 
\end{aligned}
\end{equation*}
\begin{equation*}
\begin{aligned}
\mathbb{P}^{N+1} & : [ \tau_{\mathcal{M}(1)} + 2 \Omega, \tau_{\mathcal{M}(2)} + \Omega, \cdots, \tau_{\mathcal{M}(N)} + \Omega ]^T, \\
\vdots \\
\mathbb{P}^{2^B N} & : [ \tau_{\mathcal{M}(1)}, \tau_{\mathcal{M}(2)}, \cdots, \tau_{\mathcal{M}(N)} ]^T.
\end{aligned}
\end{equation*}

Further utilize expressions $\omega_{\mathcal{M}(n),\text{opt}} = \mathbb{P}(n) - \tau_{\mathcal{M}(n)}$ and $\mathcal{M}^{-1}$, we can finally obtain the search set $\mathbb{U} = 
\left \{ \{ \omega^1_{1,\text{opt}}, \ldots, \omega^1_{N,\text{opt}} \}, \cdots, \{ \omega^{2^B N}_{1,\text{opt}}, \ldots, \omega^{2^B N}_{N,\text{opt}} \} \right \}$.
Then the original problem (P3) is reduced to 
$
\max_{\omega_1, \dots, \omega_{N} \in \mathcal{U}} \ \left| \mathbf{w}^H \boldsymbol{\phi} \right| = \max_{ \{ \omega_1, \dots, \omega_{N} \} \in {\mathbb{U}}} \ \left| \mathbf{w}^H \boldsymbol{\phi} \right|
$. 
We summarize the algorithm for case of rank-1 in Algorithm~\ref{alg2}.

\begin{algorithm}
\caption{DaS algorithm for rank-1} 
\label{alg2} 
\begin{algorithmic}[1] 
\REQUIRE Complex vector $\boldsymbol{\phi}$, the phase quantization level $B$. 
\ENSURE Optimal discrete phase configurations  $\omega_{1,\text{opt}}, \ldots, \omega_{N,\text{opt}}.$ 
\STATE For each $0 \le n \le N$, calculate $\tau_n = \text{angle} (\phi_n)$. 
\STATE Divide the complete angle into $2^B$ uniformly parts based on each $\tau_n$ and construct the center set $\mathcal{C}^n$. 
\STATE Sort $\{ \tau_1, \ldots, \tau_N \}$ and find the mapping $\mathcal{M}$.
\STATE Based on the coder associated with each part, build the set $\mathbb{U}$ consisting of $2^B N$ candidates. 
\STATE Find $\omega_{1, \text{opt}}, \ldots, \omega_{N, \text{opt}} = \arg \max_{ \{ \omega_1, \dots, \omega_{N} \} \in \mathbb{U} } \ \left| \mathbf{w}^H \boldsymbol{\phi} \right|$.
\RETURN $\mathbf{w}_{\text{opt}}= \left[ e^{j \omega_{1, \text{opt}} }, \cdots, e^{j \omega_{N, \text{opt}} } \right]^H$.
\end{algorithmic} 
\end{algorithm}

\section{Numerical and Experimental Results}\label{Section4}
To assess the efficacy of the proposed DaS algorithm, we conduct a dual evaluation involving both numerical simulations\footnote{\label{foot:code}The simulation results can be reproduced using code available at: 
https://github.com/RujingXiong/RIS\_Optimization.git} and real-world field trials. In the simulations, we compare the proposed algorithm's performance against several state-of-the-art methods, namely SDR-SDP, Manopt, and APX \cite{zhang2022configuring}. Furthermore, we execute field tests within an actual RIS-aided communication prototype system.

\begin{figure}[htbp]
  \centering
  \subfigure[]{
  \label{fig: Compa}
  \includegraphics[width=.479\columnwidth]{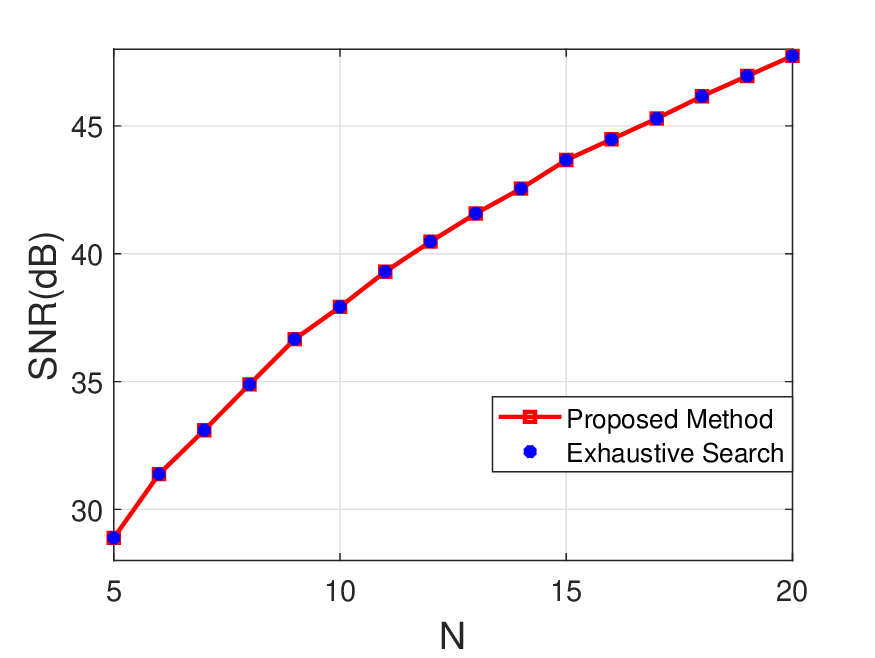}}
  \subfigure[]{
  \label{fig: Compa3Method}
  \includegraphics[width=.479\columnwidth]{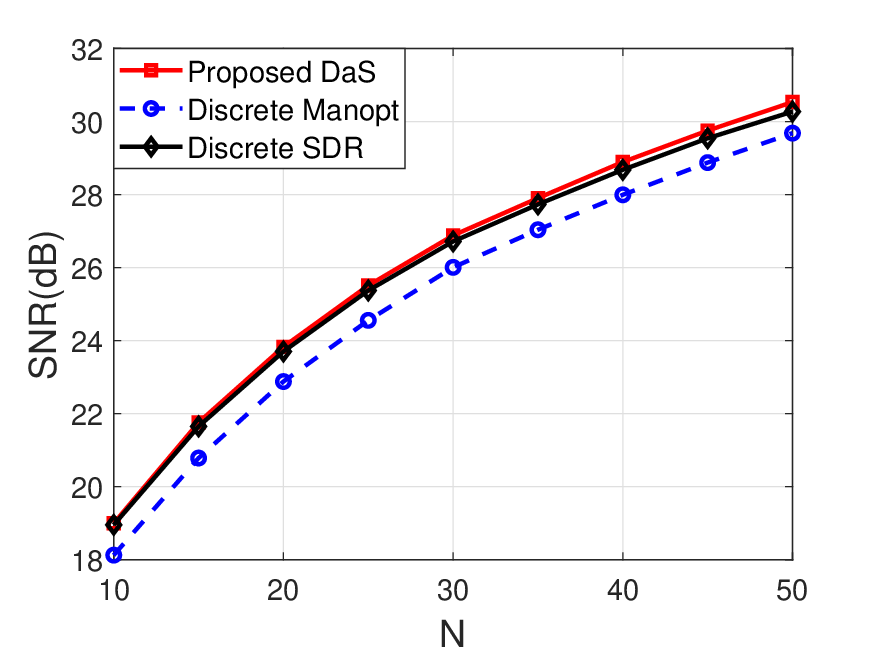}}
  \caption{A comparison of SNR performance while $M=3$. (a) The proposed DaS achieves comparable performance to that of the exhaustive search, $B=1$. (b) The SNR performance of different methods, $B=2$. }
  \label{Fig1}
\vspace{-0.1cm}
\end{figure}

%

\subsection{Signal Power Gains}
We initiate numerical simulations to evaluate the SNR performance of various methods concerning the number of units $N$. In all simulations, the equivalent channels are i.i.d. Gaussian random variables of mean zero and variance $c$. 
We refer to the SDR-SDP and Manopt algorithms, which take continuous solutions and discretization processes as discrete SDR-SDP and discrete Manopt, respectively.

To examine the optimality of the proposed DaS algorithm, we compare the average SNR achieved by DaS with that obtained through brute-force searches in low-rank ($M=3$) scenarios, as described in~\eqref{P5}. The average power gain is recorded over 400 trials with each $N$. As shown in Fig.~\ref{fig: Compa}, the plots demonstrate that DaS can achieve exactly identical SNR results to the exhaustive search method.

In Fig.~\ref{fig: Compa3Method}, we present a performance comparison with several practical methods. The results show that when using the 2-bit quantization scheme, the proposed DaS algorithm achieves an average SNR gain of 0.3 dB and 1 dB over discrete SDR-SDP and discrete Manopt, respectively.


\begin{figure}[htbp]
  \centering
  \subfigure[]{
  \label{1-bit}
    \includegraphics[width=.479\columnwidth]{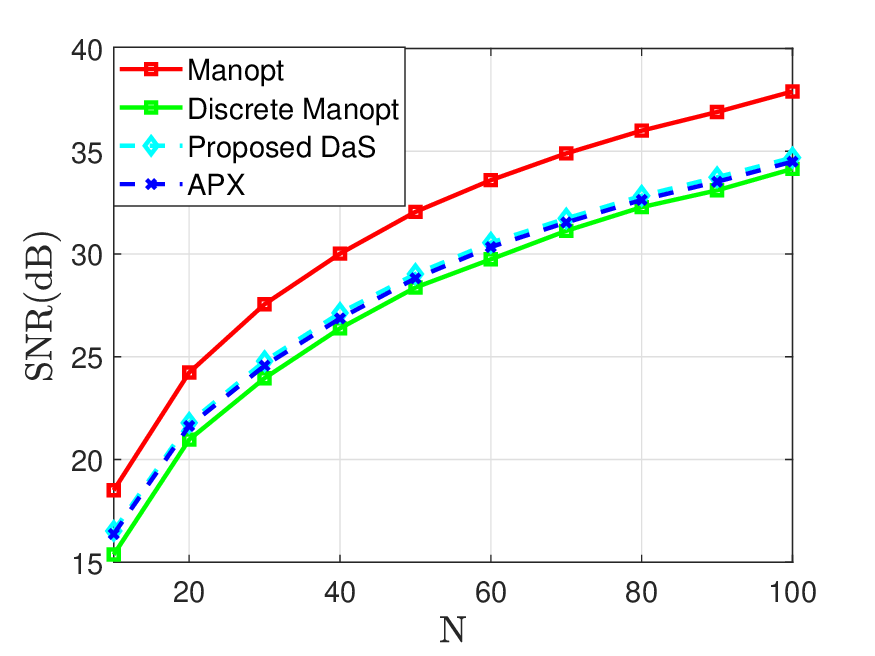}}
  \subfigure[]{
  \label{2-bit}
  \includegraphics[width=.479\columnwidth]{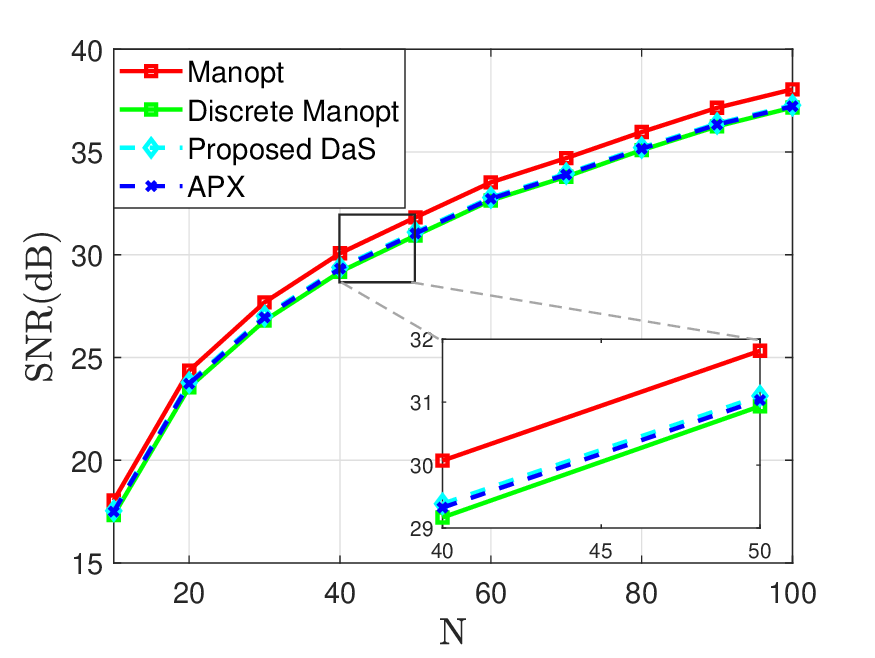}}
  \subfigure[]{
  \label{4-bit}
  \includegraphics[width=.479\columnwidth]{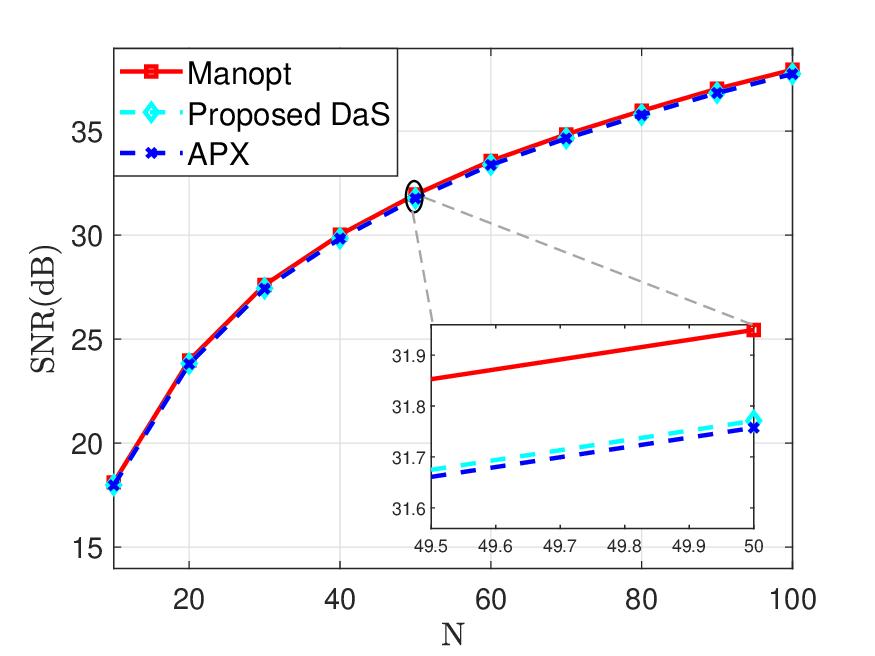}}
  \subfigure[]{
  \label{8-bit}
  \includegraphics[width=.479\columnwidth]{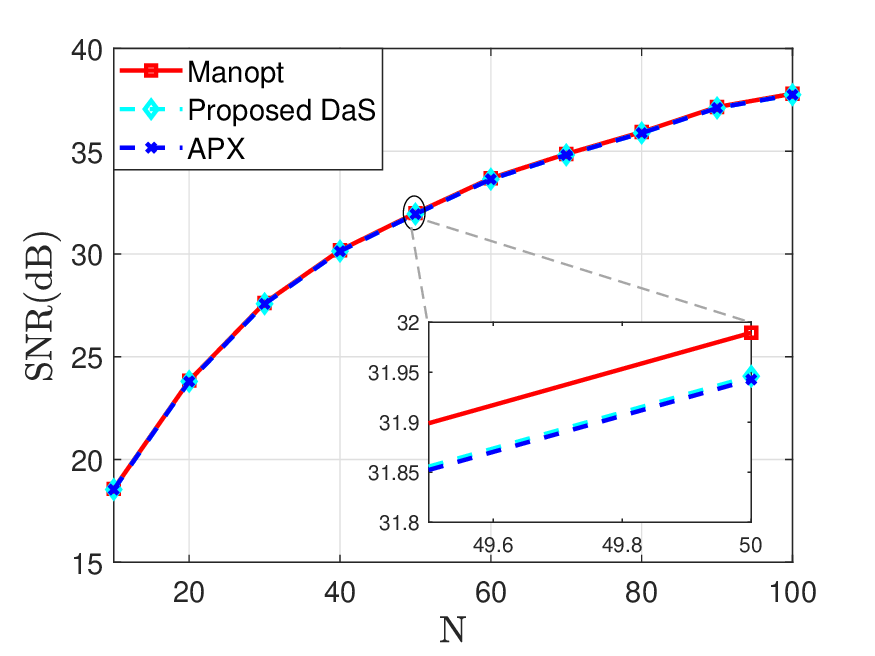}}
  \caption{SNR performance for different methods as a function of $N$, rank-1. There appears to be no significantly noticeable difference in power gains between continuous and discrete phase configurations for moderate resolution quantizations, e.g., 4-bit and above. (a) 1-bit. (b) 2-bit. (c) 3-bit. (d) 4-bit.}
  \label{2-4-8bit}
\vspace{-0.2cm}
\end{figure}


\begin{table}
\caption{Overall execution-time comparison in 1-bit RIS (rank-1)} 
\centering
\setlength{\tabcolsep}{-0.1mm}
\begin{tabular}{c|c|c|c|c|c|c|c}
Methods           &  $N = 10$  & $N = 50$    & $N = 100$  &$N = 200$   &$N = 500$   & $N=1000$ \\\hline
Exhaustive~Search & $0.96$~s    & -          & -          & -          & -          &  - \\
Discrete~SDR-SDP  &$169.26$~s   &$191.67$~s  &$312.45$~s  & $903.77$~s  &$10785.55$~s & -                         \\
Discrete~Manopt   & $3.26$~s    & $26.11$~s  &$48.50$~s   & $80.95$~s   &$260.90$~s   & $950.24$~s            \\
APX               &$0.13$~s     &$0.24$~s    &$0.47$~s    &$1.28$~s     & $8.51$~s    &$66.81$~s \\
Proposed~DaS      & $0.03$~s    & $0.09$~s   &$0.17$~s    & $0.48$~s    &$3.59$~s     & $18.81$~s  
\label{tab: time}
\vspace{-0.3cm}
\end{tabular}
\end{table}

We demonstrate that the 4-bit phase configuration is adequate in practice. In Fig.~\ref{2-4-8bit}, we include the results of continuous phase configurations (obtained using Manopt) depicted with red curves to provide a comprehensive analysis. 
It is observed that the 1-bit discrete configurations lead to a loss of about 3 dB compared to the continuous phase configurations (Manopt). However, as the quantization resolution increases, the loss in received signal power diminishes. Remarkably, when utilizing 4-bit quantization, the proposed DaS method achieves SNR gains that are comparable to continuous configurations, with an average loss of less than 0.05 dB.




\begin{figure}[htbp]
\centering
\includegraphics[width=0.55\columnwidth]{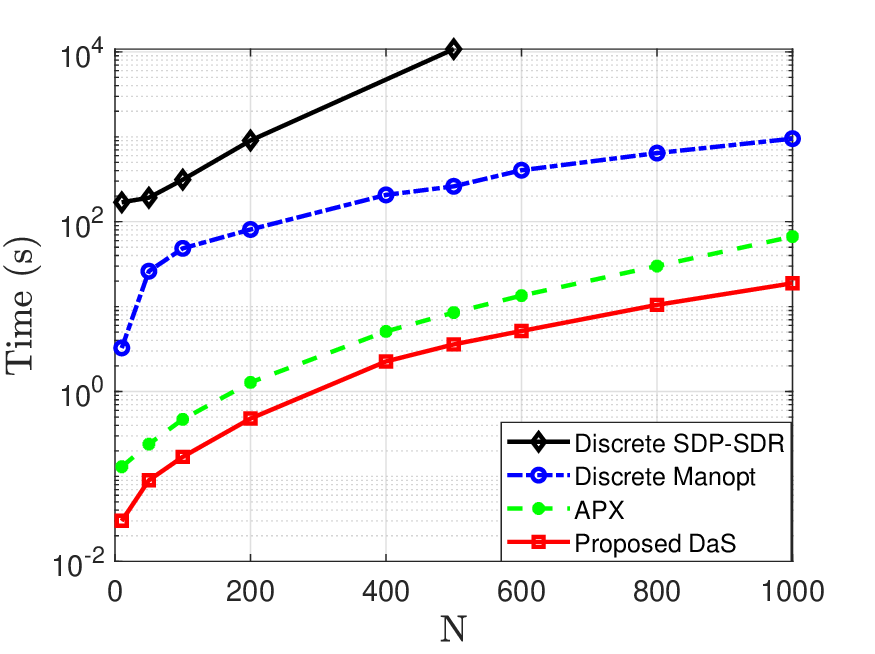}
\caption{Overall execution-time of $100$ trials for each $N$ in 1-bit RIS, rank-1.}
\label{fig:time}
\end{figure}

\subsection{Overall Execution-time Comparison}
To assess the effectiveness of the proposed approach, we conduct tests to evaluate the execution time as a function of the number of reflecting units $N$. For each $N$, we conduct 100 trials and record the total execution time as presented in Table~\ref{tab: time} and Fig.~\ref{fig:time}. The results demonstrate that the execution time of discrete SDR-SDP increases substantially as $N$ increases. In fact, for $N = 1000$, it fails to produce a solution even after running for several hours. In contrast, the proposed DaS method is much more efficient, when $N = 1000$, it takes only 0.1881 s on average to find the optimal discrete phase configurations. 
The superior execution-time performance of DaS makes it particularly well-suited for practical implementations.

%

\subsection{Field Trial Results}
\begin{figure*}[htbp]
\centering
\subfigure[]{
  \label{system}
  \includegraphics[width=0.459\linewidth]{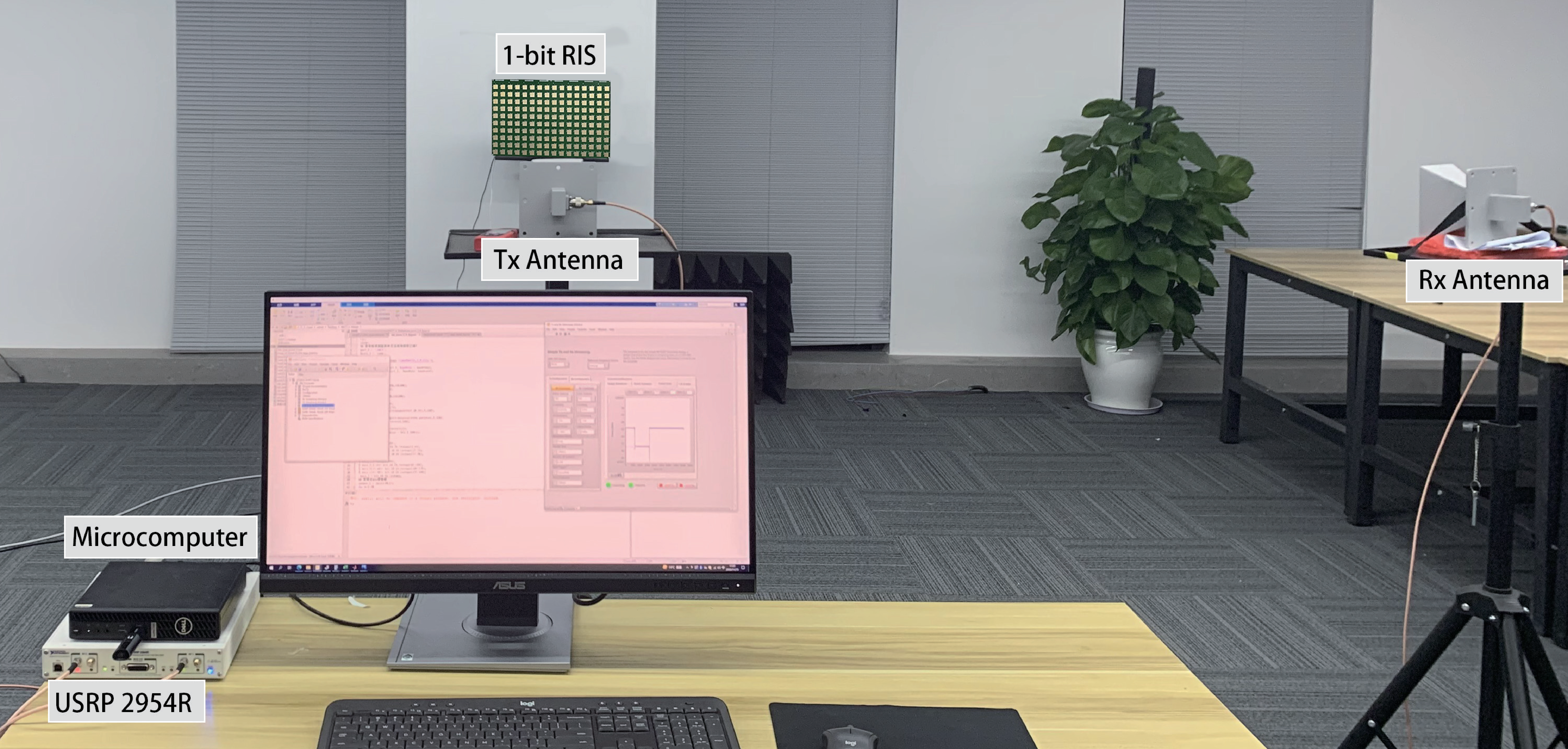}}
 \subfigure[]{
  \label{Set45}
  \includegraphics[width=.34085\linewidth]{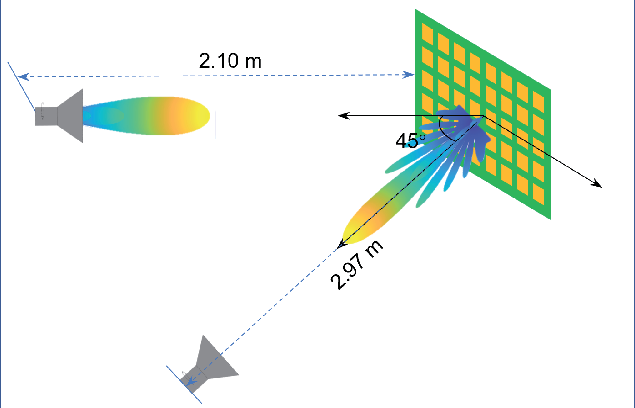}}
  \caption{Experimental setup. (a) The prototype of a 1-bit RIS-aided communication system. (b) The transmitter is in front of the RIS, and the receiver is positioned at the same height as the transmitter, with an angle of $45^{\circ}$ from the normal of the RIS.}
\label{Experiment Test}
\vspace{-0.1cm}
\end{figure*}

The objective of the prototype experiment is to compare the actual signal power gains achieved by various algorithms in an open office area, as illustrated in Fig.~\ref{system}. The signal is generated and modulated using the USRP 2954R, transmitted through a horn antenna, reflected by the 1-bit RIS operating at 5.8 GHz, and finally received by another horn. To ensure the reliability of the evaluation, we calculate the average received signal power based on 8912 samples.

In the experiment, the transmitting and receiving antennas are placed at distances of 2.1 and 2.97 meters, respectively, with elevation angles $0^{\circ}$ and $45^{\circ}$ from the RIS. All devices are situated at the same height.

\begin{figure}
\centering
\includegraphics[width=0.98\columnwidth]{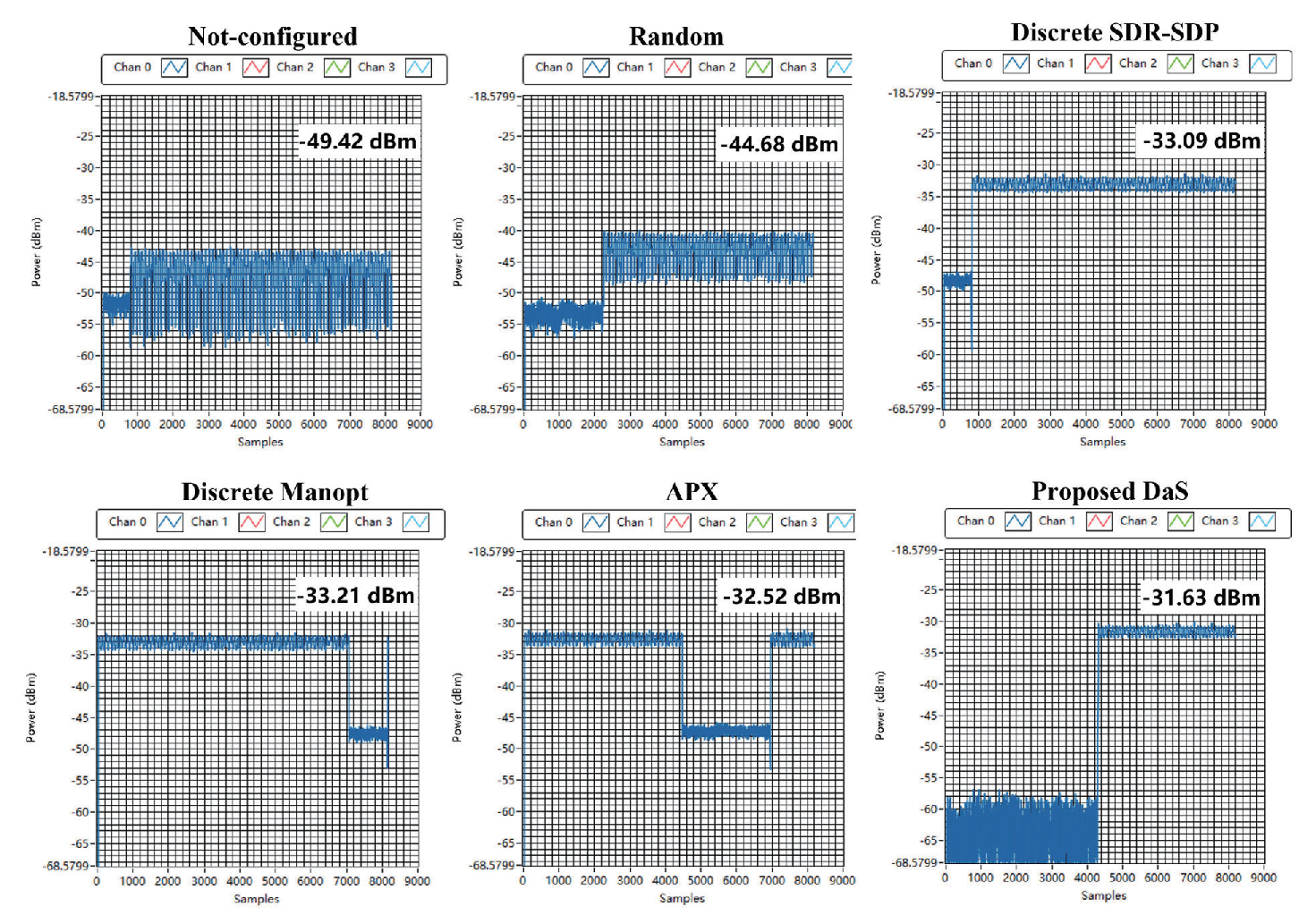}
\caption{The received signal power measurements for different approaches.}
\label{Result2.1-45}
\vspace{-0.1cm}
\end{figure}

The measurement results presented in Fig.~\ref{Result2.1-45} clearly demonstrate the effectiveness of the proposed DaS method, surpassing other competing methods by approximately 1 dB regarding received signal power gain. Notably, when compared to no elaborate phase configurations, the proposed DaS method achieves a remarkable power gain of up to 18 dB. 
 
\section{Conclusion}\label{Section5}
This paper introduces a novel framework focused on maximizing inner products to address the discrete beamforming problem in RIS-aided communication systems. The proposed DaS algorithm is tailored for the effective handling of such challenges. It is worth noting that this algorithm boasts polynomial search complexity and assures a globally optimal solution. The numerical simulations demonstrate its superiority in terms of SNR performance and execution-time efficiency when compared to alternative optimization-based approaches. Additionally, field trial results provide further validation of the efficacy of the proposed DaS algorithm in real-world scenarios, showcasing power gains of up to 18 dB over non-configured setups and surpassing other methods. Furthermore, the research findings emphasize that 4-bit discrete phase configurations can attain SNR performance on par with that of continuous configurations.

\bibliographystyle{IEEEtran}
\bibliography{Reference}

\begin{thebibliography}{10}
\providecommand{\url}[1]{#1}
\csname url@samestyle\endcsname
\providecommand{\newblock}{\relax}
\providecommand{\bibinfo}[2]{#2}
\providecommand{\BIBentrySTDinterwordspacing}{\spaceskip=0pt\relax}
\providecommand{\BIBentryALTinterwordstretchfactor}{4}
\providecommand{\BIBentryALTinterwordspacing}{\spaceskip=\fontdimen2\font plus
\BIBentryALTinterwordstretchfactor\fontdimen3\font minus
  \fontdimen4\font\relax}
\providecommand{\BIBforeignlanguage}[2]{{%
\expandafter\ifx\csname l@#1\endcsname\relax
\typeout{** WARNING: IEEEtran.bst: No hyphenation pattern has been}%
\typeout{** loaded for the language `#1'. Using the pattern for}%
\typeout{** the default language instead.}%
\else
\language=\csname l@#1\endcsname
\fi
#2}}
\providecommand{\BIBdecl}{\relax}
\BIBdecl

\bibitem{basar2019wireless}
E.~Basar, M.~Di~Renzo \emph{et~al.}, ``Wireless communications through
  reconfigurable intelligent surfaces,'' \emph{IEEE Access}, vol.~7, pp.
  116\,753--116\,773, 2019.

\bibitem{xiong2023ris}
R.~Xiong, J.~Zhang \emph{et~al.}, ``{RIS}-aided wireless communication in
  real-world: Antennas design, prototyping, beam reshape and field trials,''
  \emph{arXiv preprint arXiv:2303.03287}, 2023.

\bibitem{arun2020rfocus}
V.~Arun and H.~Balakrishnan, ``{RFocus}: Beamforming using thousands of passive
  antennas.'' in \emph{NSDI}, 2020, pp. 1047--1061.

\bibitem{pei2021ris}
X.~Pei, H.~Yin \emph{et~al.}, ``{RIS}-aided wireless communications:
  Prototyping, adaptive beamforming, and indoor/outdoor field trials,''
  \emph{IEEE Transactions on Communications}, vol.~69, no.~12, pp. 8627--8640,
  2021.

\bibitem{dai2020reconfigurable}
L.~Dai, B.~Wang \emph{et~al.}, ``Reconfigurable intelligent surface-based
  wireless communications: Antenna design, prototyping, and experimental
  results,'' \emph{IEEE Access}, vol.~8, pp. 45\,913--45\,923, 2020.

\bibitem{rains2021high}
J.~Rains, A.~Tukmanov \emph{et~al.}, ``High-resolution programmable scattering
  for wireless coverage enhancement: An indoor field trial campaign,''
  \emph{arXiv preprint arXiv:2112.11194}, 2021.

\bibitem{wu2019intelligent}
Q.~Wu and R.~Zhang, ``Intelligent reflecting surface enhanced wireless network
  via joint active and passive beamforming,'' \emph{IEEE Transactions on
  Wireless Communications}, vol.~18, no.~11, pp. 5394--5409, 2019.

\bibitem{wu2019beamforming}
------, ``Beamforming optimization for wireless network aided by intelligent
  reflecting surface with discrete phase shifts,'' \emph{IEEE Transactions on
  Communications}, vol.~68, no.~3, pp. 1838--1851, 2019.

\bibitem{wang2022sca}
T.~Wang, F.~Fang \emph{et~al.}, ``An {SCA} and relaxation based energy
  efficiency optimization for multi-user ris-assisted {NOMA} networks,''
  \emph{IEEE Transactions on Vehicular Technology}, vol.~71, no.~6, pp.
  6843--6847, 2022.

\bibitem{kumar2022novel}
V.~Kumar, R.~Zhang \emph{et~al.}, ``A novel {SCA}-based method for beamforming
  optimization in {IRS/RIS}-assisted {MU-MISO} downlink,'' \emph{IEEE Wireless
  Communications Letters}, vol.~12, no.~2, pp. 297--301, 2022.

\bibitem{cui2019secure}
M.~Cui, G.~Zhang \emph{et~al.}, ``Secure wireless communication via intelligent
  reflecting surface,'' \emph{IEEE Wireless Communications Letters}, vol.~8,
  no.~5, pp. 1410--1414, 2019.

\bibitem{zhou2020robust}
G.~Zhou, C.~Pan \emph{et~al.}, ``Robust beamforming design for intelligent
  reflecting surface aided {MISO} communication systems,'' \emph{IEEE Wireless
  Communications Letters}, vol.~9, no.~10, pp. 1658--1662, 2020.

\bibitem{yu2019miso}
X.~Yu, D.~Xu \emph{et~al.}, ``{MISO} wireless communication systems via
  intelligent reflecting surfaces,'' in \emph{2019 IEEE/CIC International
  Conference on Communications in China (ICCC)}.\hskip 1em plus 0.5em minus
  0.4em\relax IEEE, 2019, pp. 735--740.

\bibitem{elmossallamy2021ris}
M.~A. ElMossallamy, K.~G. Seddik \emph{et~al.}, ``{RIS} optimization on the
  complex circle manifold for interference mitigation in interference
  channels,'' \emph{IEEE Transactions on Vehicular Technology}, vol.~70, no.~6,
  pp. 6184--6189, 2021.

\bibitem{wang2020intelligent}
P.~Wang, J.~Fang \emph{et~al.}, ``Intelligent reflecting surface-assisted
  millimeter wave communications: Joint active and passive precoding design,''
  \emph{IEEE Transactions on Vehicular Technology}, vol.~69, no.~12, pp.
  14\,960--14\,973, 2020.

\bibitem{zhang2022configuring}
Y.~Zhang, K.~Shen \emph{et~al.}, ``Configuring intelligent reflecting surface
  with performance guarantees: Optimal beamforming,'' \emph{IEEE Journal of
  Selected Topics in Signal Processing}, 2022.

\bibitem{di2020hybrid}
B.~Di, H.~Zhang \emph{et~al.}, ``Hybrid beamforming for reconfigurable
  intelligent surface based multi-user communications: Achievable rates with
  limited discrete phase shifts,'' \emph{IEEE Journal on Selected Areas in
  Communications}, vol.~38, no.~8, pp. 1809--1822, 2020.

\bibitem{ren2022linear}
S.~Ren, K.~Shen \emph{et~al.}, ``A linear time algorithm for the optimal
  discrete irs beamforming,'' \emph{IEEE Wireless Communications Letters},
  vol.~12, no.~3, pp. 496--500, 2022.

\bibitem{tse2005fundamentals}
D.~Tse and P.~Viswanath, \emph{Fundamentals of wireless communication}.\hskip
  1em plus 0.5em minus 0.4em\relax Cambridge University Press, 2005.

\end{thebibliography}

\end{document}